# Title: Emergence of electric-field-tunable interfacial ferromagnetism in 2D antiferromagnet heterostructures

**Authors:** Guanghui Cheng[1,2,3,12], Mohammad Mushfiqur Rahman[4], Zhiping He[5], Andres Llacsahuanga Allcca[2,3,6], Avinash Rustagi[4], Kirstine Aggerbeck Stampe[7], Yanglin Zhu[8], Shaohua Yan[9], Shangjie Tian[9], Zhiqiang Mao[8], Hechang Lei[9], Kenji Watanabe[10], Takashi Taniguchi[11], Pramey Upadhyaya[3,4,6], Yong P. Chen[1,2,3,4,6,7]*

**Affiliations:**

[1]WPI-AIMR International Research Center for Materials Sciences, Tohoku University, Sendai 980-8577, Japan.

[2]Department of Physics and Astronomy, and Birck Nanotechnology Center, Purdue University, West Lafayette, Indiana 47907, USA.

[3]Purdue Quantum Science and Engineering Institute, Purdue University, West Lafayette, Indiana 47907, USA.

[4]School of Electrical and Computer Engineering, Purdue University, West Lafayette, Indiana 47907, USA.

[5]Department of Physics, University of Science and Technology of China, Hefei, Anhui 230026, China.

[6]Quantum Science Center, Oak Ridge, Tennessee 37831 USA.

[7]Institute of Physics and Astronomy and Villum Centers for Dirac Materials and for Hybrid Quantum Materials, Aarhus University, 8000 Aarhus-C, Denmark.

[8]Department of Physics, Pennsylvania State University, University Park, Pennsylvania 16802, USA.

[9]Department of Physics and Beijing Key Laboratory of Opto-electronic Functional Materials & Micro-nano Devices, Renmin University of China, Beijing 100872, China.

[10]Research Center for Functional Materials, National Institute for Materials Science, 1-1 Namiki, Tsukuba 305-0044, Japan.

[11]International Center for Materials Nanoarchitectonics, National Institute for Materials Science, 1-1 Namiki, Tsukuba 305-0044, Japan.

[12]Current address: Department of Physics, University of Science and Technology of China, Hefei, Anhui 230026, China.

*Corresponding author. Email: yongchen@purdue.edu

**Abstract:** Van der Waals (vdW) magnet heterostructures have emerged as new platforms to explore exotic magnetic orders and quantum phenomena. Here, we study heterostructures of layered antiferromagnets, $CrI_3$ and $CrCl_3$, with perpendicular and in-plane magnetic anisotropy, respectively. Using magneto-optical Kerr effect microscopy, we demonstrate out-of-plane magnetic order in the $CrCl_3$ layer proximal to $CrI_3$, with ferromagnetic interfacial coupling between the two. Such an interlayer exchange field leads to higher critical temperature than that of either $CrI_3$ or $CrCl_3$ alone. We further demonstrate significant electric-field control of the coercivity, attributed to the naturally broken structural inversion symmetry of the heterostructure allowing unprecedented direct coupling between electric field and interfacial magnetism. These findings illustrate the opportunity to explore exotic magnetic phases and engineer spintronic devices in vdW heterostructures.

## Introduction

Heterostructures are promising to host emergent phenomena and device functions not present in constituent parts [1-10]. One well-known example is the integration of two insulating complex oxides leading to a conducting two-dimensional electron gas at the interface [2], with surprising coexistence of superconductivity and ferromagnetism [3]. The recently explored van der Waals (vdW) magnets have pushed the research frontier to 2D magnetism where exotic magnetic ground states and quantum phases can emerge [11-18]. Magnetic vdW heterostructures provide a new toolbox to explore magnetic proximity and related effects [4-10]. A largely unexplored arena is to combine two different magnetic orders and investigate the magnetic proximity at the interface, which could allow modulation of magnetic interactions and establish exotic magnetic properties. It is also of fundamental significance to effectively control the exchange interactions and magnetic anisotropy, with the latter being crucial to stabilize the long-range magnetic orders.

The studies of layered semiconducting chromium trihalides have shown exotic magnetic behaviors and rich tunability by stimuli [14-18]. Typically, the few-layer $CrI_3$ is an antiferromagnet with Ising-like perpendicular magnetic anisotropy (PMA) [19, 20], as schematically depicted in Fig. 1a. The interlayer antiferromagnetic coupling is ascribed to the exchange interactions between Cr mediated by ligand atoms [17, 20]. In contrast, few-layer $CrCl_3$ is an easy-plane antiferromagnet, where spins prefer to lie in the layers [20, 21]. In particular, the single-ion anisotropy from spin-orbit coupling (SOC) of Cr and the anisotropic exchange from SOC of Cl nearly cancel out each other [20, 21]. Therefore, $CrCl_3$ is located close to the boundary between PMA and in-plane anisotropy, suggesting that its magnetic properties may be particularly susceptible to external perturbations. The combined heterostructure of $CrI_3$ and $CrCl_3$ is possibly a fertile system to realize rich magnetic phases and manipulate them. Such a heterostructure has not yet been explored.

Here, we fabricate $CrI_3$/$CrCl_3$ heterostructures and demonstrate interfacial ferromagnetism between the two antiferromagnets. Figure 1a left panel schematically depicts the expected spin configurations for the magnetic ground states in bilayer (2L) $CrI_3$ and few-layer (FL) $CrCl_3$ with PMA and in-plane anisotropy [19-21], respectively. Due to the strong intralayer ferromagnetic coupling in chromium trihalides [17], we can denote all spins in a given layer by a macroscopic spin (out-of-plane: ↑, ↓; in-plane:←, →). The optical micrograph of a representative 2L $CrI_3$/FL $CrCl_3$ heterostructure is shown in Fig. 1b. The 2L $CrI_3$ is partially stacked on top of FL $CrCl_3$, allowing the comparison between regions of 2L $CrI_3$, FL $CrCl_3$ and heterostructure. Atomic force microscopy confirms flake thickness of 1.6 nm for 2L $CrI_3$ (Fig. 1c) and 9.5 nm for FL $CrCl_3$, respectively.

## Results and Discussion

We employ magneto-optical Kerr effect microscopy (MOKE) under the polar configuration as the primary measurement due to its high sensitivity to the magnetic moments perpendicular to the sample surface [22]. Figure 1d shows the MOKE signal ($\theta_K$) of the 2L $CrI_3$ and the 2L $CrI_3$/FL $CrCl_3$ heterostructure as a function of the perpendicular magnetic field. In the 2L $CrI_3$ region, $\theta_K$ stays close to zero at low field, corresponding to the antiferromagnetic states ↓↑ or ↑↓ with zero net magnetization. Beyond critical field ±0.76 T, $\theta_K$ abruptly jumps to ferromagnetic states with finite magnetization. This is consistent with the reported spin-flip transitions in 2L $CrI_3$ [23].

In the 2L $CrI_3$/FL $CrCl_3$ heterostructure, the antiferromagnetic-to-ferromagnetic spin-flip transition of 2L $CrI_3$ is still present and its critical field decreases from ±0.76 T in 2L $CrI_3$ region to ±0.57 T in the heterostructure region. Remarkably, a significant square hysteresis loop is observed with coercive field ~±0.1 T, indicating a magnetic transition between two different phases with non-zero net magnetization, in sharp contrast to the antiferromagnetic ground states in 2L $CrI_3$ (↓↑ or ↑↓). Such a ferromagnetic-like loop is absent in either 2L $CrI_3$ or FL $CrCl_3$, suggesting its origin from interfacial magnetic interaction. This phenomenon should not be due to the charge transfer/doping-induced antiferromagnetic-to-ferromagnetic transition reported in 2L $CrI_3$ [14, 15], which does not exhibit such a coexistence of antiferromagnetic-type and ferromagnetic-type transitions. Recent works have shown that a twist of two chromium trihalides layers may lead to noncollinear antiferromagnetic-ferromagnetic domains and thus finite MOKE signals[9, 10, 18, 24]. However, this scenario is also less likely to be relevant, as such domains are predicted to emerge for sufficiently large moiré periodicity. Due to large lattice constant mismatch [18], the $CrI_3$/$CrCl_3$ heterostructure can hardly form large moiré periodicity even at zero twist angle. Furthermore, the magnetization behaviors (including the field and temperature dependence) measured in twisted bilayer $CrI_3$ [9, 10] exhibit qualitative difference from what we observe in our samples. We further rule out several other possible origins (see detailed discussions in Supplementary Text 1). We propose that at least three spin layers (two layers of $CrI_3$ plus one neighboring layer of $CrCl_3$) are responsible for the observed transitions. The neighboring $CrCl_3$ layer is acting as the third spin layer with out-of-plane magnetic order after being stacked in proximity with $CrI_3$, as schematically shown in Fig. 1a and denoted in the red dashed rectangles in Fig. 1d,e. Note that a perpendicular magnetic field induces canting of the planar $CrCl_3$ spins, giving rise to a continuously varying MOKE background [25] (Supplementary Fig. 1), which is typically subtracted and eliminated from our MOKE signal. Therefore, only perpendicular spin-flip transitions are discussed in this work.

Similar to trilayer $CrI_3$, in principle several potential antiferromagnetic configurations can be considered: ↑↓↓ (-1), ↓↑↑ (+1), ↓↑↓ (-1), ↑↓↑ (+1) with the first two and the third spins referring to the 2L $CrI_3$ and the neighboring $CrCl_3$ layer respectively and the numbers in brackets denoting the net magnetic moments. To figure out the coupling type for the neighboring $CrI_3$ and $CrCl_3$ layers, we study the 1L $CrI_3$/FL $CrCl_3$ heterostructure where the magnetic behavior can directly verify the interlayer coupling type. Figure 1e shows $\theta_K$ of the 1L $CrI_3$ and the 1L $CrI_3$/FL $CrCl_3$ heterostructure, fabricated from the same $CrI_3$ flake. Interestingly, both show a ferromagnetic behavior with a single hysteresis loop. This observation indicates that the neighboring $CrI_3$ and $CrCl_3$ layer is ferromagnetically coupled, in contrast to the interlayer antiferromagnetic coupling in FL $CrI_3$ [19]. Careful inspection on the hysteresis loop in Fig. 1d shows more transition steps, possibly due to the switching of magnetic domains [23, 26]. Note that due to thin-film optical interference effect [23, 27]. it's not possible to associate the magnitude or sign of the MOKE signal to the magnetization of different samples (e.g., the doubling of the MOKE signal in the 1L $CrI_3$/FL $CrCl_3$ heterostructure relative to that in the 1L $CrI_3$ in Fig. 1e does not imply that the probed magnetization doubles). In the following text, we mainly focus on other features (e.g., emergent hysteresis loop, transition/coercive fields, critical temperatures). To exclude any unique causes related to the stacking sequence, we also studied reversely stacked heterostructures with FL $CrCl_3$ on top of 2L $CrI_3$ and observed similar hysteresis loops (Supplementary Figs. 2 and 3). The larger coercive field of the hysteresis loop observed in the reversed stack may be due to sample differences or twist angle dependence and is out of the scope of this work.

We next study the temperature dependence of the magnetism in the heterostructure. Figure 2a,b shows the temperature dependence of $\theta_K$ in 2L $CrI_3$ and 2L $CrI_3$/FL $CrCl_3$ heterostructure. The extracted $H_1$, $H_1^*$, $H_2^*$ and $\Delta\theta_1$, $\Delta\theta_1^*$, $\Delta\theta_2^*$ as a function of temperature are shown in Fig. 2c,d, respectively. The antiferromagnetic spin-flip transitions (at $H_1$ and $H_1^*$) in both 2L $CrI_3$ and the heterostructure disappear at temperatures larger than $T_C \sim 40$ K and is consistent with previous measurements in 2L $CrI_3$ [28]. The ferromagnetic-like hysteresis loop (at $H_2^*$) observed only in the heterostructure region survives up to a higher temperature $T_C^* \sim 48$ K. Another experiment on 1L $CrI_3$/FL $CrCl_3$ heterostructure shows critical temperatures of $T_C \sim 33$ K and $T_C^* \sim 37$ K for 1L $CrI_3$ region and heterostructure region, respectively (Supplementary Fig. 4). In 2D magnets, the critical temperature is determined by the spin-wave excitation gap, which is dictated by the anisotropies present in the system [12, 23, 29, 30]. Our density functional theory (DFT) results suggest an increase in the effective single-ion anisotropy of $CrI_3$ when brought in proximity to $CrCl_3$. On the other hand, thanks to the induced ferromagnetic coupling, the $CrCl_3$ layer now sees an effective anisotropy field that depends both on the interfacial ferromagnetic coupling as well as the anisotropy of $CrI_3$, which is expected to enlarge the spin-wave gaps for both the materials. This is consistent with the observed increase of $T_C$ for both 1L $CrI_3$/FL $CrCl_3$ and 2L $CrI_3$/FL $CrCl_3$ systems.

To better understand the observations, we explore the magnetic ground states of the $CrI_3$/$CrCl_3$ bilayer using first-principles calculations (Supplementary Text 2). We find that the perpendicular ferromagnetic state (↑↑) is more favorable than three other magnetic configurations: perpendicular antiferromagnetic state (↑↓), the states that one layer is out-of-plane polarized while the other in-plane polarized (↑→ or →↑). Further consideration on magnetic dipole-dipole interaction and different commensurate twist angles (0˚ and 30˚) does not undermine the favorable perpendicular ferromagnetic state. The interlayer exchange energy $J_{\text{inter}}$ in $CrI_3/CrCl_3$ can be approximated by the energy difference between perpendicular antiferromagnetic and ferromagnetic configurations [17]. We estimate $J_{\text{inter}} \approx$ -77 (-64) μJ/m$^2$ for 0˚ (30˚)-twisted $CrI_3/CrCl_3$, compared to the reported interlayer exchange ~80 μJ/m$^2$ in 2L $CrI_3$ [29, 31, 32]. Such interfacial exchange coupling in the heterostructure wins over the in-plane anisotropy of $CrCl_3$ and results in the out-of-plane magnetic order in the $CrCl_3$ layer next to $CrI_3$, in agreement with our observations.

We next turn to explore the electrical tunability of the observed interfacial magnetism. A unique aspect of the $CrI_3/CrCl_3$ heterostructure, when compared with previously explored monolayer and/or homobilayer systems [14, 15, 33], is the absence of structural inversion symmetry. The Neumann's principle [34] states that the spin-charge coupling is dictated by the symmetries of the system. We thus expect to observe spin-charge coupling phenomena for the interlayer magnetic order. In particular, breaking of structural inversion allows for direct electric-field modification of the magnetic anisotropy and the interlayer exchange interactions via terms of the form (see detailed discussions in Supplementary Text 3):

$$E_{\text{elec}}(\mathbf{m}_i, \sigma_i) = (\sigma_1 - \sigma_2)(\beta_1 m_{z1}^2 + \beta_2 m_{z2}^2 + \beta_3 \mathbf{m}_1 \cdot \mathbf{m}_2), \qquad (1)$$

where $\mathbf{m}_i$, $\sigma_i$ are the magnetization and charges of the respective layers, $(\sigma_1 - \sigma_2) \sim$ electric field and $\beta_{1,2,3}$ parameterizes the strength of respective interactions. Microscopically, the electric-field control of interfacial magnetic interactions could arise from electric-field-induced changes in the orbital occupancy in conjunction with spin-orbit interactions. Such a mechanism has attracted significant interest for constructing low-dissipation spintronic memory and logic devices [35, 36].

To check the electric-field tuning of the observed interfacial magnetism, we fabricated a dual-gated 1L $CrI_3$/FL $CrCl_3$ device, as shown in Fig. 3a,b. This structure allows us to study the magnetization of the $CrI_3/CrCl_3$ heterostructure (as well as that of the 1L $CrI_3$ region in the same device) under the top-gate voltage $V_{\text{tg}}$ and back-gate voltage $V_{\text{bg}}$. The two voltages are converted to electrostatic doping density $n$ and displacement field $D$ (Methods). Figure 3c shows the coercive field ($H_c$) in 1L $CrI_3$/FL $CrCl_3$ heterostructure increases from ~700 Oe to ~1000 Oe when the $D$ is tuned from -1.4 V $nm^{-1}$ to 1 V $nm^{-1}$, indicating the enhancement of the magnetic anisotropy of the interfacial ferromagnetism in the heterostructure. The full mappings in Fig. 3d,e present the extracted $H_c$ as a function of both $n$ and $D$ in 1L $CrI_3$ and the 1L $CrI_3$/FL $CrCl_3$ heterostructure, respectively. A quite weak modulation of $H_c$ is observed in 1L $CrI_3$, suggesting that the magnetism of 1L $CrI_3$ can hardly be tuned under the range of gating voltages of this work. Separate experiments on FL $CrCl_3$ demonstrate that the magnetism of $CrCl_3$ also can hardly be tuned by electrostatic gating (Supplementary Fig. 5d). However, significant tunability of the $H_c$ is observed by the $D$ applied to the heterostructure. Such a dramatic tunability in the $CrI_3$/$CrCl_3$ is in agreement with the electric field control of interfacial magnetic interactions allowed by the structural symmetry breaking, predicted in the above theoretical analysis. The intriguing electrical tunability allowed by symmetry breaking is also observed in a heterostructure containing a bilayer $CrI_3$ (Supplementary Text 4).

In summary, we studied the interfacial magnetism in $CrI_3$/$CrCl_3$ heterostructures and demonstrated the interfacial ferromagnetic coupling between neighboring $CrI_3$ and $CrCl_3$ layers. The demonstrated ability to engineer magnetoelectric phenomena by breaking symmetries via vdW heterostructures provides opportunities for vdW spintronics. The compatible hybrids of 2D magnets with other quantum materials, such as unconventional superconductors, ferroelectrics or topological materials are predicted to demonstrate exotic topological phases and many-body interactions[12, 13, 37], as well as to design new spintronic devices and therefore are highly desirable for further study.

## Methods

**Crystal growth.** Single crystal $CrI_3$ was synthesized using the chemical vapor transport (CVT) method [38]. The Cr powder and iodine pieces were mixed with a stoichiometric ratio and loaded into a quartz tube (inner diameter, 10 mm; length, 180 mm). The quartz tube was sealed under vacuum and then transferred to a double temperature zones furnace. The temperatures of the hot and cold ends of the furnace were set at 650 °C and 550 °C, respectively. The growth with such a temperature gradient lasted for 7 days. Finally, the furnace was shut down, and the quartz tube naturally cooled down to room temperature. The black plate-like $CrI_3$ crystals can be found at the cold end of the quartz tube.

Single crystal $CrCl_3$ was grown by the CVT method. The commercial $CrCl_3$ polycrystal powder (99.9%) was sealed in a silica tube with a length of 200 mm and an inner diameter of 14 mm. The tube was pumped down to 0.01 Pa and sealed under vacuum, and then placed in a two-zone horizontal tube furnace. The two growth zones were raised slowly to 973 K and 823 K for 2 days, and then held there for another 7 days. After that, the furnace was shut down and cooled down naturally. Shiny, plate-like crystals with lateral dimensions of up to several millimeters can be obtained from the growth.

**Device fabrication.** Few-layer $CrI_3$, $CrCl_3$ and hexagonal boron nitride (hBN) flakes are exfoliated onto the silicon wafer covered by 285-nm thermal oxide layer. Flakes with proper thickness are selected by optical contrast [23] and later confirmed by atomic force microscopy (AFM) and MOKE measurements. $CrI_3$ flakes used in this work have 1~2 layers and the few-layer $CrCl_3$ flakes are around 5~10 nm (0.6 nm for each layer) thick. Heterostructures of $CrI_3$ and $CrCl_3$ are fabricated by the dry-transfer method and encapsulated between two hBN flakes with a typical thickness of ~10 nm. Specifically, a stamp made of a thin polycarbonate and polydimethylsiloxane is then employed to pick up the flakes in sequence under an optical microscope. In the end, the finished stack is deposited onto the target substrate with polycarbonate on top which is removed by chloroform afterwards. The whole process is performed inside a glovebox to avoid material degradation. The exposure time to air is kept below ten minutes before transferring the fabricated sample into the measurement chamber and pumping down.

For the dual-gated heterostructure device and magnetic tunneling junction device, few-layer graphene flakes are exfoliated and integrated into the stack following the above processes. The target substrate is pre-patterned with electrodes fabricated by standard e-beam lithography, Au/Ti deposition and lift-off processes. The stack is carefully aligned and transferred onto the target pattern to make contact between graphene flakes and electrodes.

**MOKE microscopy.** The polarization of a linearly polarized light reflected from a magnetic material will be rotated by a Kerr angle $\theta_K$, which is proportional to the magnetization of the material. In this work, the incident light is normal to the sample plane and MOKE is in the polar geometry, meaning that the magnetic vector being probed is perpendicular to the sample surface and parallel to the incident light. A balanced photodetector and lock-in method are used to obtain the MOKE signal. A laser is used here with wavelength of 633 nm and power of 5 μW. The sample is placed in a helium-flow optical cryostat with the temperature down to 6 K and magnetic field (perpendicular to sample surface) up to 5 T. The laser is focused onto the sample surface by an objective with the spot diameter of 0.5 μm.

**Electrical control of the dual-gated device.** Top-gate and back-gate voltages can be applied to the few-layer graphene gates in the heterostructure device, while the graphene contact to the heterostructure is grounded. The dual-gate structure allows independent control of the doping density and displacement field applied on the heterostructure. The doping density $n$ and displacement field $D$ are extracted by the simple parallel plate capacitor model. For simplicity, the $CrI_3/CrCl_3$ heterostructure is regarded as one channel, on which the doping density and electric field are applied. The quantum capacitance of $CrI_3$ and $CrCl_3$ is much larger than that of graphene due to the nearly flat bands of these two magnetic semiconductors [15]. Therefore, only geometric capacitances $C_{\mathrm{bg}}$ and $C_{\mathrm{tg}}$ are considered. The doping density and displacement field can be written as $n = C_{\mathrm{bg}} \cdot V_{\mathrm{bg}} + C_{\mathrm{tg}} \cdot V_{\mathrm{tg}}$ and $D = (D_{\mathrm{bg}} + D_{\mathrm{tg}})/2 = (\varepsilon_{\mathrm{bg}} \cdot V_{\mathrm{bg}}/d_{\mathrm{bg}} - \varepsilon_{\mathrm{tg}} \cdot V_{\mathrm{tg}}/d_{\mathrm{tg}})/2$, respectively. The relative dielectric constant of hBN [14] is $\varepsilon_{\mathrm{bg}} = \varepsilon_{\mathrm{tg}} = 3$. For the device in Fig. 3, the thicknesses of bottom hBN and top hBN are obtained by AFM measurement to be $d_{\mathrm{bg}} = 19.6$ nm and $d_{\mathrm{tg}} = 14.9$ nm, respectively.

**Data and code availability**. The data and code supporting the findings of this study are included in the paper and its Supplementary Information file. Further data and code sets are available from the corresponding author on reasonable request.

## Acknowledgements

We thank Di Xiao, Wenguang Zhu for helpful discussions and Adam W. Tsen for help with crystals. The first-principles calculations have been done on the supercomputing system in the Supercomputing Center of the University of Science and Technology of China. **Funding:** We acknowledge partial support of the work from WPI-AIMR, JSPS KAKENHI Basic Science A (18H03858), New Science (18H04473 and 20H04623), Tohoku University FRiD program, Department of Defense (DOD) Multidisciplinary University Research Initiatives (MURI) program (FA9550-20-1-0322), US Department of Energy (DOE) Office of Science through the Quantum Science Center (QSC, a National Quantum Information Science Research Center), and Villum Foundation. Z.Q.M. acknowledges the support by the US DOE under grants DE-SC0019068 for sample synthesis. H.C.L. acknowledges the support by National Key R&D Program of China (2018YFE0202600), Beijing Natural Science Foundation (Z200005), and National Natural

Science Foundation of China (11822412 and 11774423). K.W. and T.T. acknowledge support from the Elemental Strategy Initiative conducted by the MEXT, Japan (JPMXP0112101001) and JSPS KAKENHI (19H05790, 20H00354 and 21H05233).

**Author contributions**

G.H.C. and Y.P.C. conceived the project. G.H.C. fabricated the devices and performed experiments, assisted by A.L.A. M.M.R., Z.P.H., A.R., K.A.S. and P.U. performed supporting theoretical modeling. Y.L.Z. and Z.Q.M. provided bulk $CrI_3$ crystals. S.H.Y, S.J.T. and H.C.L. provided bulk $CrCl_3$ crystals. K.W. and T.T. provided bulk hBN crystals. Y.P.C. supervised the project. G.H.C., M.M.R., Z.P.H., P.U. and Y.P.C. wrote the manuscript with input from all authors.

**Competing interests**

The authors declare no competing interests.

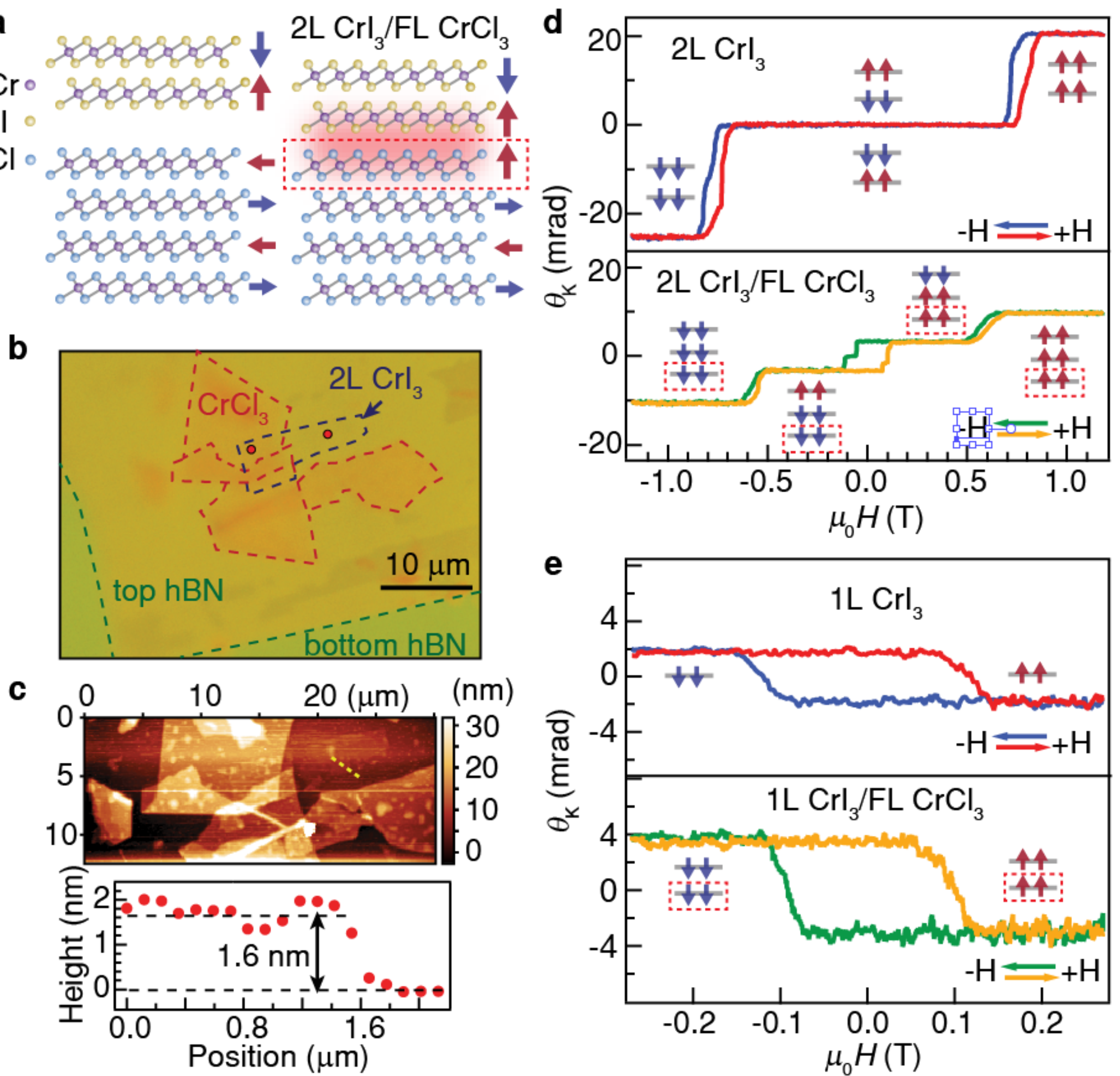

**Figure 1 | $CrI_3$/$CrCl_3$ heterostructures and MOKE measurements. a,** Schematics of the magnetic ground states in bilayer (2L) $CrI_3$ and few-layer (FL) $CrCl_3$ before (left) and after (right) forming heterostructure. Only four layers of $CrCl_3$ are shown for simplicity. **b,** Optical micrograph of a 2L $CrI_3$/ FL $CrCl_3$ heterostructure. **c,** Atomic force microscopy of the heterostructure in the same position as in **b**. The height profile (along the yellow dotted line in the image) at the edge of $CrI_3$ indicates the thickness of a bilayer. **d,** MOKE signal (after subtracting a polynomial background, as done for all MOKE curves in the main text) of the 2L $CrI_3$ region and the 2L $CrI_3$/FL $CrCl_3$ heterostructure region as a function of perpendicular magnetic field. Two curves of each region represent forward and backward sweeps of the field, respectively. The data is taken at the spots marked by red in **b**. Insets depict magnetic ground states of 2L $CrI_3$ and the $CrI_3$/$CrCl_3$ heterostructure (showing only the interfacial $CrCl_3$ layer, highlighted by red dashed rectangles). **e,** MOKE signal of another monolayer (1L) $CrI_3$/FL $CrCl_3$ heterostructure, compared with that measured in the 1L $CrI_3$ (from the same $CrI_3$ flake as in the heterostructure region).

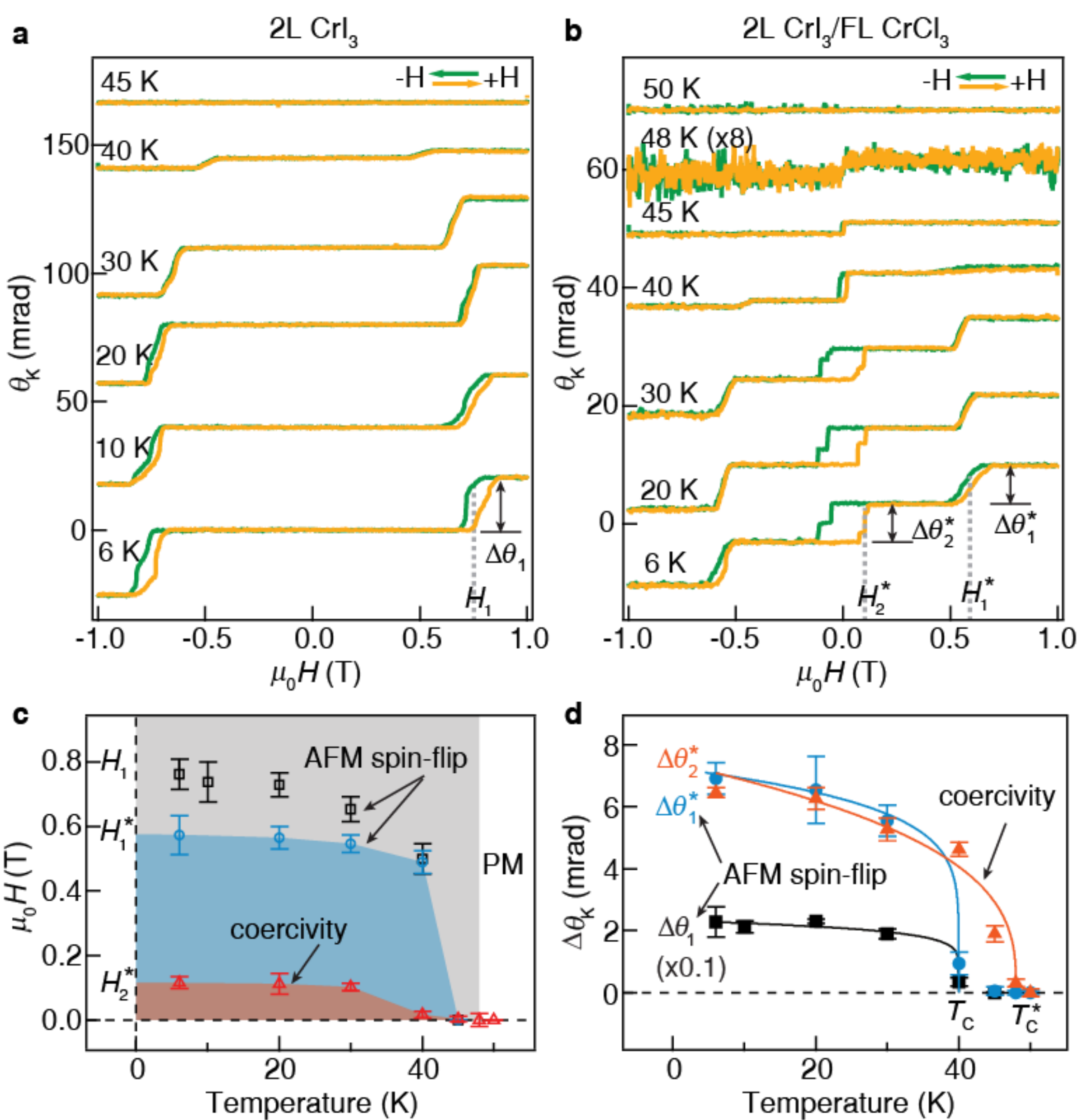

**Figure 2 | Temperature dependence of the magnetism of 2L $CrI_3$ and 2L $CrI_3$/FL $CrCl_3$ heterostructure. a,b,** MOKE signal in the 2L $CrI_3$ region (**a**) and the 2L $CrI_3$/FL $CrCl_3$ heterostructure region (**b**) as a function of perpendicular magnetic field at different temperatures. Critical fields $H_1$, $H_1^*$, $H_2^*$ and magnitudes in the change of MOKE signal $\Delta\theta_1$, $\Delta\theta_1^*$, $\Delta\theta_2^*$ of magnetic transitions are labeled. **c,** Temperature dependence of the critical fields of magnetic transitions. PM: paramagnetic. The critical fields are extracted from the peak of derivative $d\theta_K/dH$ and the error bars are the peak widths. **d,** Temperature dependence of the magnitudes in the change of MOKE signal at magnetic transitions. Solid curves are fitted by a power-law equation [11]. The critical temperatures $T_C$ and $T_C^*$ are indicated. The error bars are the uncertainties in extracting the transition magnitudes.

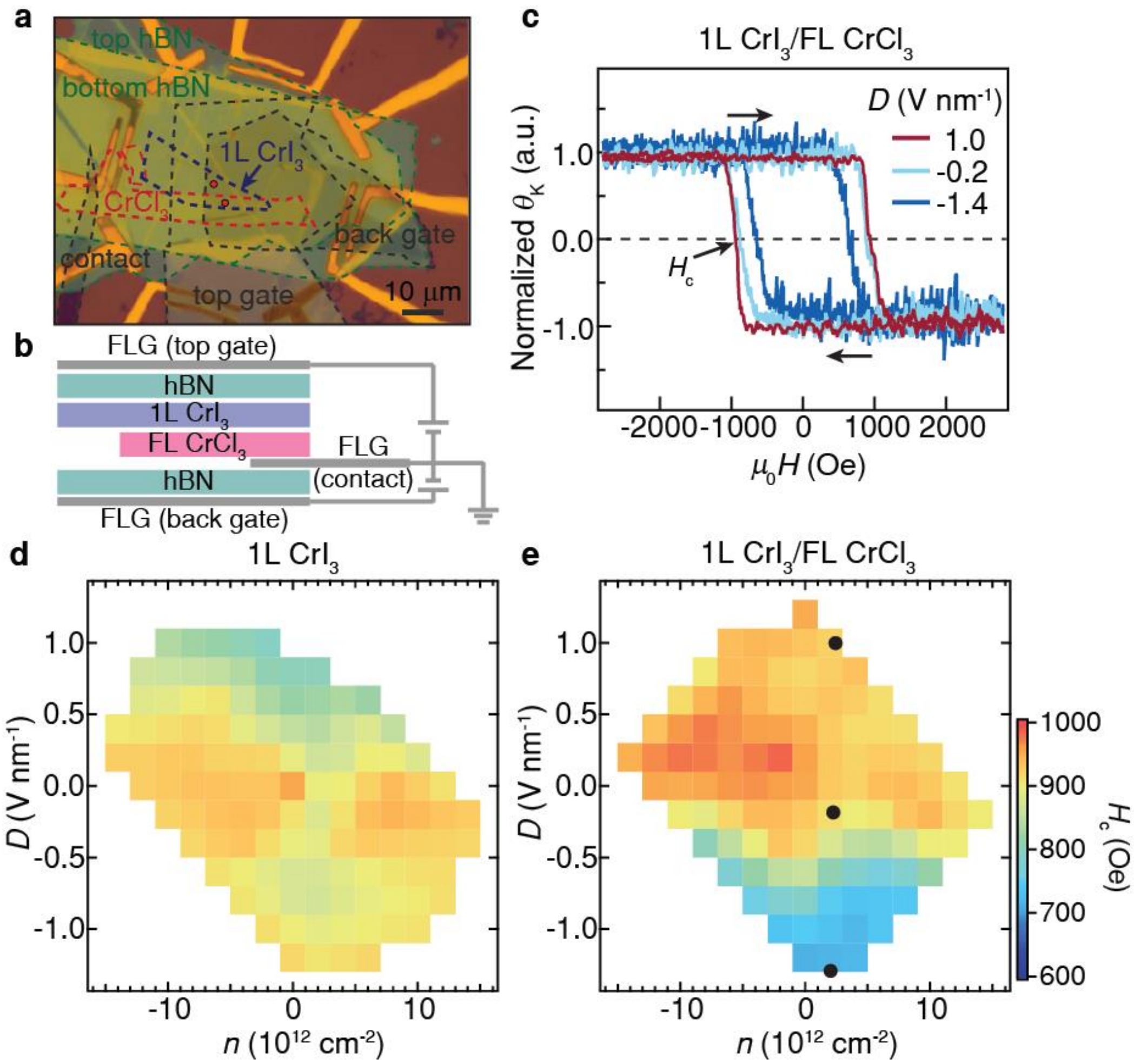

**Figure 3 | Electrical control of the magnetism in 1L $CrI_3$/FL $CrCl_3$ heterostructure. a,b,** Optical micrograph and schematic structure of a dual-gated 1L $CrI_3$/FL $CrCl_3$ device. Three few-layer graphene (FLG) flakes are used as back/top gates and the contact to the stack. **c,** Normalized MOKE signal as a function of perpendicular magnetic field in 1L $CrI_3$/FL $CrCl_3$ heterostructure under different displacement fields $D$ = -1.4, -0.2, 1.0 V nm$^{-1}$. **d,e,** Coercive field $H_c$ as a function of electrostatic doping density $n$ and displacement field $D$ of 1L $CrI_3$ (**d**) and 1L $CrI_3$/FL $CrCl_3$ heterostructure (**e**), respectively. The black dots in **e** correspond to the MOKE curves in **c**. The data is taken at the spots marked by red in **a**.